\documentclass[conference]{IEEEtran}

\usepackage{graphicx}
\usepackage{epstopdf}
\usepackage{multirow}
\usepackage{diagbox}
\usepackage{amsmath}
\usepackage{listings}
\usepackage[ruled, linesnumbered]{algorithm2e}
\usepackage{algorithmic}
\usepackage[bookmarks=false]{hyperref}
\usepackage{lineno}
\usepackage{rotating}
\usepackage{color}
\usepackage{setspace}
\usepackage{float}
\usepackage{subcaption}
\begin{document}
%
\title{Evaluating Security and Availability of Multiple Redundancy Designs when Applying Security Patches}

\author{\IEEEauthorblockN{Mengmeng Ge}
\IEEEauthorblockA{University of Canterbury\\
Christchurch, New Zealand\\
Email: mge43@uclive.ac.nz}
\and
\IEEEauthorblockN{Huy Kang Kim}
\IEEEauthorblockA{Korea University\\
Seoul, South Korea\\
Email: cenda@korea.ac.kr}
\and
\IEEEauthorblockN{Dong Seong Kim}
\IEEEauthorblockA{University of Canterbury\\
Christchurch, New Zealand\\
Email: dongseong.kim@canterbury.ac.nz}}

\maketitle

\begin{abstract}
In most of modern enterprise systems, redundancy configuration is often considered to provide availability during the part of such systems is being patched. However, the redundancy may increase the attack surface of the system. In this paper, we model and assess the security and capacity oriented availability of multiple server redundancy designs when applying security patches to the servers. We construct (1) a graphical security model to evaluate the security under potential attacks before and after applying patches, (2) a stochastic reward net model to assess the capacity oriented availability of the system with a patch schedule. We present our approach based on case study and model-based evaluation for multiple design choices. The results show redundancy designs increase capacity oriented availability but decrease security when applying security patches. We define functions that compare values of security metrics and capacity oriented availability with the chosen upper/lower bounds to find design choices that satisfy both security and availability requirements.
\end{abstract}

\IEEEpeerreviewmaketitle
\section{Introduction}
Nowadays, most enterprises have a centralized patch management system in order to install vendors’ patches efficiently and enhance their system’s security level. A security patch is an update to fix vulnerabilities in the software to prevent systems from possible exploits. However, some critical security patches require system’s reboot. When a server is under security patch, it cannot provide normal service and introduce downtime. In order to improve the availability, most enterprises have redundant servers with high availability configuration (e.g., active-active or active-standby). Ironically, this kind of redundancy may increase the attack surface because of the increased number of servers. An attack surface of a system refers to the total number of vulnerabilities that are accessible to an attacker~\cite{Manadhata2011SE}. From this view, an attacker has more number of vulnerabilities when redundancy provision is used. Besides, it is not feasible to patch all vulnerabilities of the servers due to time and cost constraints. Therefore, it is important to find the balance between the security and availability affected by the security patch to facilitate the better redundancy design for enterprise systems.

Graphical security models (e.g., attack graphs (AGs)~\cite{Sheyner2002SP}, attack trees (ATs)~\cite{Saini2008JCSC}) have been widely used to assess the cyber security. In particular, an AG captures all possible sequences of an attacker's actions to compromise the target, and an AT explores possible ways that how an attack goal is achieved via combinations of attacks. In order to improve the scalability problem of AGs and ATs, the multi-layer hierarchical attack representation model (HARM)~\cite{Hong2016JNCA} was proposed to combine AGs and ATs. In the two-layered HARM, the upper layer represents the network reachability information (i.e., nodes connected in the topological structure) and the lower layer denotes the vulnerability information of nodes, respectively.

Stochastic models have been applied to assess the availability. Stochastic Reward Net (SRN) was developed as a modeling formalism for the automatic generation and solution of the underlying continuous time Markov chain (CTMC). SRNs can be automatically constructed and converted into CTMC models using software packages such as SHARPE~\cite{Trived2009SHARPE} and SPNP~\cite{Ciardo1989SPNP}.

In this work, we construct a graphical security model (we used a HARM) to analyze the security of enterprise networks under potential attacks before and after patch, and a stochastic model (we used a SRN) to assess the system availability with patch schedule. To the best of our knowledge, this work is the first approach to evaluate both security and capacity oriented availability of redundancy designs under the security patch. The main contributions of this paper are summarized as follows:
\begin{itemize}
\item Use model-based evaluation to investigate the impact of security patch on security and capacity oriented availability of enterprise networks; 
\item Compare the security and capacity oriented availability of multiple design choices for server redundancy;
\item Find the design choices which satisfy both the security and availability requirements.
\end{itemize}

The rest of the paper is organized as follows. Section~\ref{relatedwork} presents related work for modeling both security and dependability. Our proposed approach is described in Section~\ref{approach}. Numerical analysis using the proposed approach is presented in Section~\ref{analysis}. Limitations and potential extensions are presented in Section~\ref{limit}. Finally, Section~\ref{conclusion} concludes the paper.

\section{Related Work}
\label{relatedwork}

There are extensive works on applying graphical security models in analyzing security and stochastic models assessing availability for various systems. 

\textbf{Graphical security models:} Roy \emph{et al.}~\cite{Roy2011SECN} proposed attack countermeasure trees (ACTs) for the qualitative and probabilistic security analysis by taking into account defense mechanisms on the nodes. They implemented the ACT in the SHARPE~\cite{Trived2009SHARPE} and showed the usability of their model in case study. Albanese \emph{et al.}~\cite{Albanese2012DSN} used AGs to compute the minimum-cost network hardening solution. The experiments were carried out using synthetic attack graphs and the results validated the performance of their approach. Hong \emph{et al.}~\cite{Hong2016JNCA} developed the multi-layered HARM and performed the scalability analysis compared with the single layer AGs in terms of model construction and evaluation. The simulation results demonstrated that the HARM is more scalable than the single layer AG.

\textbf{Availability models:} Kim \emph{et al.}~\cite{Kim2009PRDC} proposed a hierarchical approach to model the availability of the non-virtualized and virtualized systems using the fault tree for the system and the CTMC for the components. Trivedi \emph{et al.}~\cite{Trivedi2013ASMBI} presented several case studies on assessing the availability of real systems from Motorola, Cisco and Sun Microsystems via stochastic models.

There are a few work focused on considering both security and dependability. Trivedi \emph{et al.}~\cite{Trivedi2009DRCN} proposed a new classification of dependability and security models and showed case studies for both the individual and composite models. Wang \emph{et al.}~\cite{Wang2003OASIS} developed an intrusion tolerant architecture for distributed servers based on fault tolerant computing techniques to mitigate the impact of both known and unknown attacks. Bangalore \emph{et al.}~\cite{Bangalore2009DEPEND} proposed the method of self-cleansing intrusion tolerance based on the virtualization technique by reducing the server's exposure time to less than a minute. The experiment results showed lower exposure time leads to slightly larger response time but yields higher security level. Yu \emph{et al.}~\cite{Yu2010SECRYPT} evaluated the survivability (both statically and under sustained attacks) and costs of three virtual machine based architectures using the analytical methods. Ramasamy \emph{et al.}~\cite{Ramasamy2007DSN} used combinatorial modeling to analyze the impact of virtualization on the single physical node based on the assumption that module failures are independent.

\section{A Proposed Approach}
\label{approach}
\begin{figure}[htb]
\centering
\includegraphics[width=3.25in]{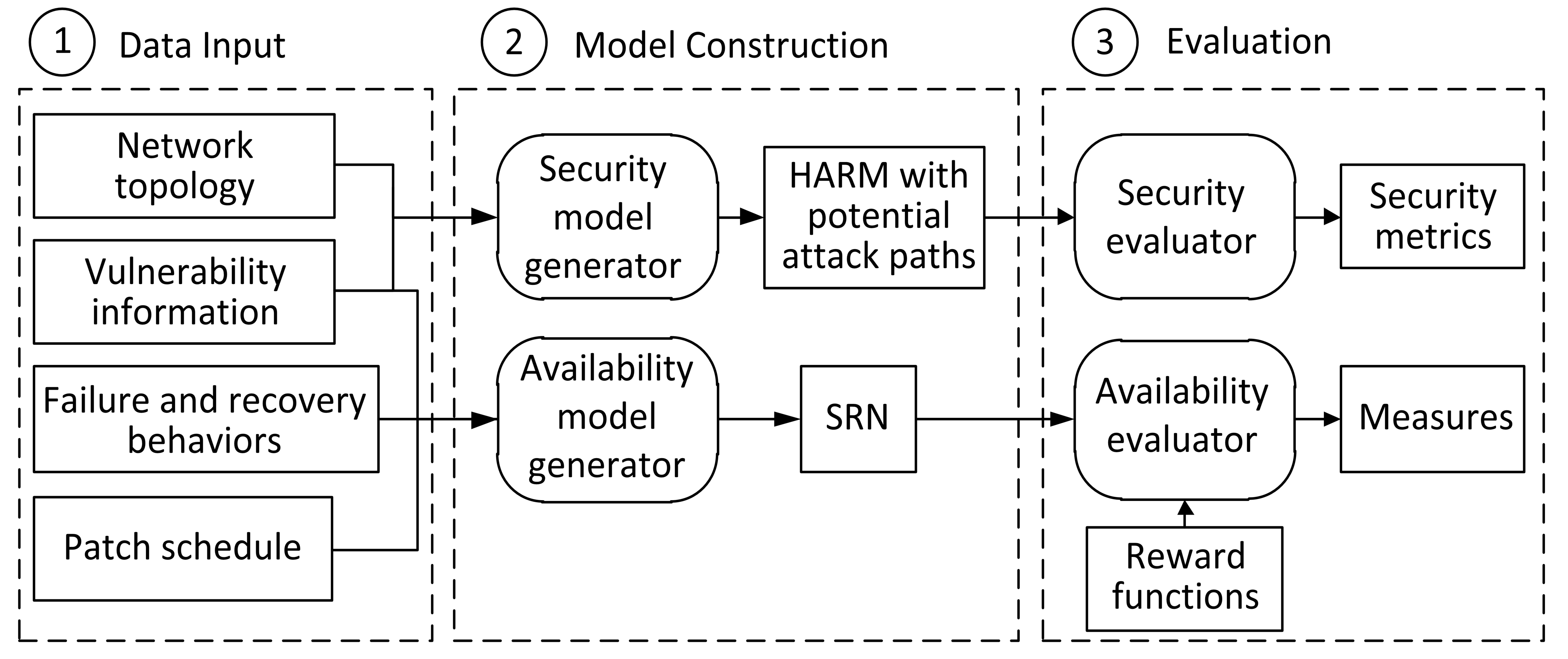}
\caption{The proposed approach.}
\label{fig_flow}
\end{figure}

The approach consists of three phases shown in Figure~\ref{fig_flow}: 1) data input, 2) model construction, and 3) evaluation. We explain each phase in the following.

In the phase 1, the administrator needs to provide four types of inputs: the network topology (i.e., reachability information), node vulnerability information, failure and recovery behaviors of the nodes in the network (i.e., interactions between nodes due to failure and recovery, failure and recovery rates), and patch schedule (i.e., how often to patch, which vulnerabilities to be patched). The vulnerability information includes the Common Vulnerabilities and Exposures (CVE) ID, the Common Vulnerability Scoring System (CVSS) base score~\cite{Gallon2011ARES} and other metric-based values assigned to the vulnerability (e.g., attack success probability, attack impact). The network topology and vulnerability information are used as inputs into the security model generator. The failure and recovery behaviors, patch schedule and the related vulnerability information are used as inputs into the availability model generator.

In the phase 2, we perform the construction of two models (security and availability) based on the inputs separately. First, the security model generator automatically generates a two-layered HARM for the network with the potential attack paths captured in the upper layer. Second, we use SPNP~\cite{Ciardo1989SPNP} as the availability model generator and manually construct the SRN model. An automated SRN model construction can be performed as in ~\cite{Machida2011Candy}. 

In the phase 3, we carry out the evaluation of security and availability. The security evaluator computes two types of security metrics: path-based (e.g., number of attack paths) and non-path-based (e.g., attack success probability). Other security analysis is also possible to perform as in ~\cite{Ge2017JNCA}. The availability evaluator outputs measures (e.g., capacity oriented availability) using the module in the SPNP which takes the pre-defined reward functions. Security metrics and availability measures can be combined for further analysis as shown in Section~\ref{comparison}.

If there are any changes about the network (e.g., new vulnerabilities are found), both models should be re-constructed with the new inputs. In this paper, we only consider one-time patch (e.g., monthly patch of a specific month) and analyze the impact of the patch in terms of security and availability. More complex cases (e.g., monthly patch of 3 months) will be considered in our future work. 

\subsection{An Example Enterprise Network}
\label{net}
We demonstrate the feasibility of our approach using case study. We present an example enterprise network along with the potential attacker model at first and then build the HARM and availability model. Our proposed approach is not limited to this specific case but applicable to general enterprise networks.

\begin{figure}[htb]
\centering
\includegraphics[width=3.25in]{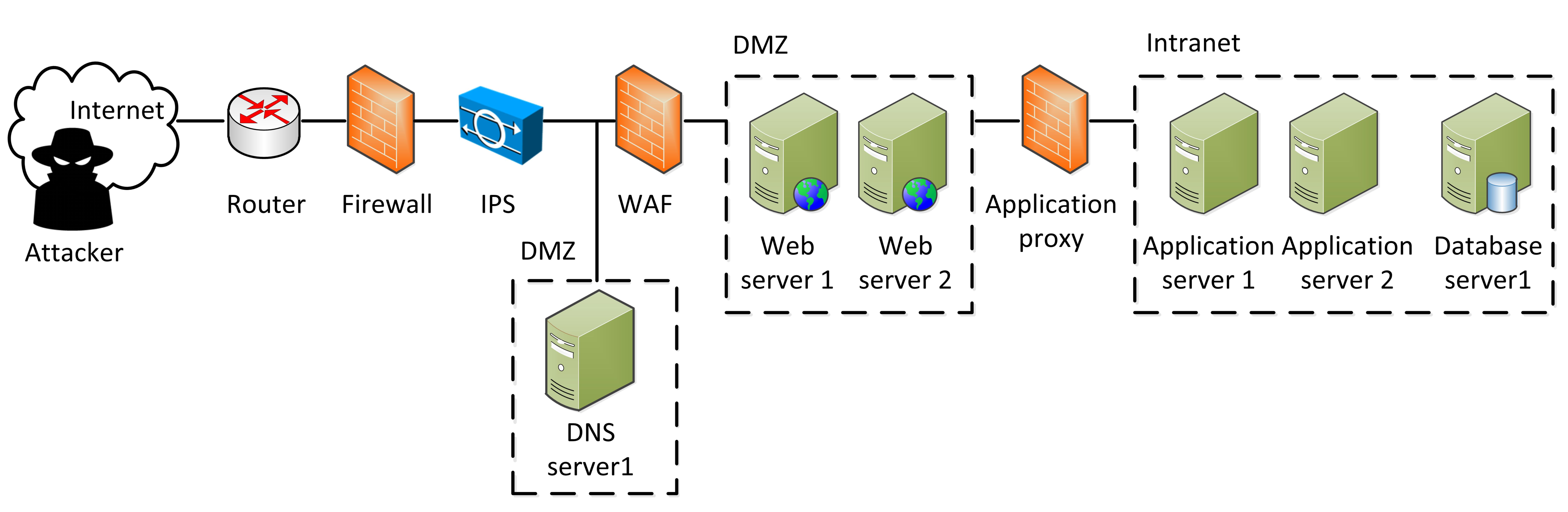}
\caption{An example enterprise network.}
\label{fig_net}
\end{figure}

We assume the enterprise network uses a 3-tier client-server architecture for its web service. The network is divided into three subnets by the external and internal firewalls shown in Figure~\ref{fig_net}. The web servers are deployed in the demilitarized zone (DMZ). The application servers and database server are located in the intranet. There is also a domain name service (DNS) server responding to the domain name queries in another DMZ. Both web servers and application servers use active-active high availability cluster configuration. The redundant servers are identical in terms of both hardware and software.

We make assumptions for the services running on the servers: the DNS server runs Windows Server 2012 R2 and Microsoft DNS; the web server runs Red Hat Enterprise Linux and Apache HTTP; the application server runs Oracle Linux 7 and Oracle WebLogic; the database server runs Oracle Linux 7 and MySQL. All the chosen operating systems (OSs) and service software are commonly used in the enterprise networks.

We collect the vulnerability information for each server from the National Vulnerability Database (NVD) and use the vulnerabilities for the example network before patch. In the real networks, vulnerabilities can be collected from scanning tools. Each vulnerability has a base score from the CVSS to indicate its severity. We denote the exploitable vulnerabilities as vulnerabilities which can be exploited by a remote attacker to gain some level of privileges. We denote critical vulnerabilities as vulnerabilities with the CVSS base score higher than 8.0. A vulnerability can be both exploitable and critical, or neither exploitable nor critical.

Each server has a patch schedule. We only consider the regular patch for security vulnerabilities and use monthly patch (30 days) for the network as patches are usually released monthly. Other patch management will be considered in our future work. As it is impossible to patch every vulnerability due to cost ant downtime, scenarios of patching critical vulnerabilities will be taken into account. 

\subsection{An Attacker Model}
\label{attack}

We make the following assumptions for the capability of an attacker as follows. The attacker is located outside the network and the attack goal is to compromise the database server(s) through privilege escalation attacks. The effort required to compromise one server has no correlation with the effort required to compromise another server (e.g., a single attack tool cannot exploit a vulnerability in the web server and another vulnerability in the application server).

\subsection{Construction of HARMs}
\label{sec}

We make use of the two-layered HARM proposed in~\cite{Hong2016JNCA}. We choose the following metrics in our security evaluation: the attack success probability, attack impact, number of exploitable vulnerabilities, number of attack paths and number of entry points~\cite{Pendleton2016ComSur}. The CVSS base score of each vulnerability is calculated from the impact and exploitability scores. We extract the value of attack success probability from the CVSS exploitability score by dividing the score by 10. We use the impact score as the value of attack impact. We present the impact and probability values for the exploitable vulnerabilities in Table~\ref{tb_vuls_cve}.
\begin{table}[htb] \small
\caption{Vulnerability information of the example network.}
\label{tb_vuls_cve}
\centering
\begin{tabular}{|c|c|c|c|}
\hline
\multirow{2}{*}{Vulnerability} & \multirow{2}{*}{CVE ID} & Attack & Attack success\\
&&impact&probability\\
\hline
$v_{1_{\mathit{dns}}}$ & CVE-2016-3227 & 10.0 & 1.0 \\
\hline
$v_{1_{\mathit{web}}}$ & CVE-2016-4448 & 10.0 & 1.0\\
\hline
$v_{2_{\mathit{web}}}$ & CVE-2015-4602 & 10.0 & 1.0\\
\hline
$v_{3_{\mathit{web}}}$ & CVE-2015-4603 & 10.0 & 1.0\\
\hline
$v_{4_{\mathit{web}}}$ & CVE-2016-4979 & 2.9 & 1.0\\
\hline
$v_{5_{\mathit{web}}}$ & CVE-2016-4805 & 10.0 & 0.39\\
\hline
$v_{1_{\mathit{app}}}$ & CVE-2016-3586 & 10.0 & 1.0\\
\hline
$v_{2_{\mathit{app}}}$ & CVE-2016-3510 & 10.0 & 1.0\\
\hline
$v_{3_{\mathit{app}}}$ & CVE-2016-3499 & 10.0 & 1.0\\
\hline
$v_{4_{\mathit{app}}}$ & CVE-2016-0638 & 6.4 & 1.0\\
\hline
$v_{5_{\mathit{app}}}$ & CVE-2016-4997 & 10.0 & 0.39\\
\hline
$v_{1_{\mathit{db}}}$ & CVE-2016-6662 & 10.0 & 1.0\\
\hline
$v_{2_{\mathit{db}}}$ & CVE-2016-0639 & 10.0 & 1.0\\
\hline
$v_{3_{\mathit{db}}}$ & CVE-2015-3152 & 2.9 & 0.86\\
\hline
$v_{4_{\mathit{db}}}$ & CVE-2016-3471 & 10.0 & 0.39\\
\hline
$v_{5_{\mathit{db}}}$ & CVE-2016-4997 & 10.0 & 0.39\\
\hline
\end{tabular}
\end{table}

\begin{figure}[hbt]
    \centering
    \begin{subfigure}{0.45\textwidth}
            \includegraphics[width=\textwidth]{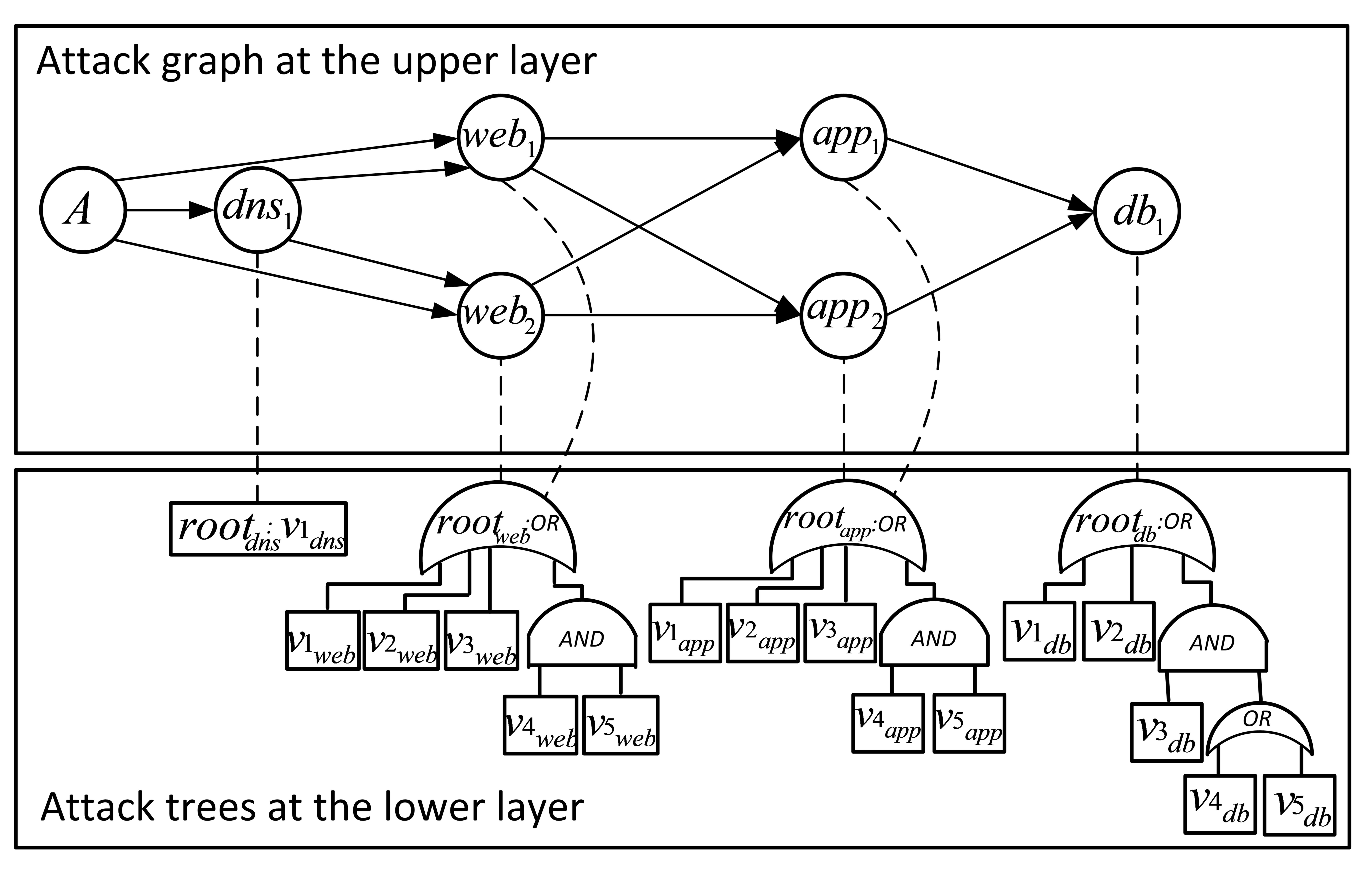}
            \caption{Before patch is complete.}
    \end{subfigure}
    \begin{subfigure}{0.275\textwidth}
            \includegraphics[width=\textwidth]{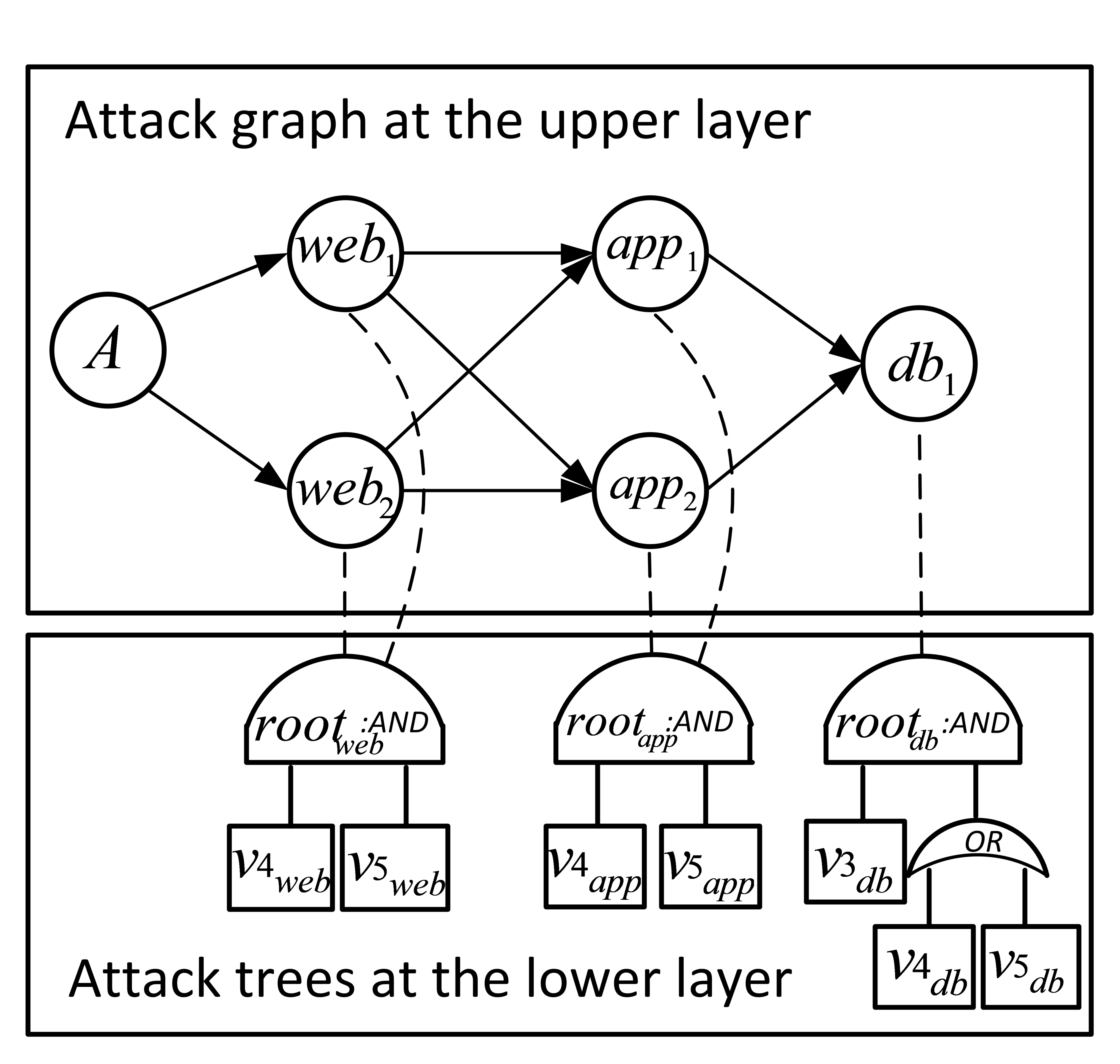}
            \caption{After patch is complete.}
    \end{subfigure}
    \caption{HARMs of the example network.}
    \label{fig_harm}
\end{figure}

We construct the two-layered HARMs for the example enterprise network before and after patch in Figure~\ref{fig_harm}(a) and~\ref{fig_harm}(b), respectively. In the HARM, the upper layer represents the network reachability information using an AG (e.g., in Figure~\ref{fig_harm}(a), an attacker $A$ is able to reach the target $\mathit{db}_1$ via $\mathit{dns}_1$, ($\mathit{web}_1$ or $\mathit{web}_2$), ($\mathit{app}_1$ or $\mathit{app}_2$)) and the lower layer denotes the vulnerability information of each server using ATs (e.g., in Figure~\ref{fig_harm}(a), an attacker $A$ is able to exploit either $v_{1_{\mathit{web}}}$, or $v_{2_{\mathit{web}}}$, or $v_{3_{\mathit{web}}}$, or both $v_{4_{\mathit{web}}}$ and $v_{5_{\mathit{web}}}$ to gain the root permission of the web server).

The number of exploitable vulnerabilities in the network is calculated by adding the number of exploitable vulnerabilities in each server. As the potential attack paths are captured in the HARM,   the number of attack paths and number of entry points are calculated in the upper layer of the HARM. For the attack impact and attack success probability, values in the higher levels are calculated from the lower levels. All the equations can be found in~\cite{Ge2017JNCA,Yusuf2016CIT,Moon2016CIT}. We show an example of calculating the attack impact in the vulnerability, node, attack path and network levels for the network before patch to demonstrate how the security analysis is carried out using the HARM.

Let $\mathit{aim}_{\mathit{root}_{\mathit{svr}}}$ denote the attack impact calculated recursively in the corresponding AT containing the vulnerability information for a server $svr$, $\mathit{aim}_{\mathit{svr}}$ denote the attack impact in the node level. We calculate $\mathit{aim}_{\mathit{root}_{\mathit{web}_1}}$ and assign the value to $\mathit{aim}_{\mathit{web}_1}$ in the following.
\begin{align*} 
\mathit{aim}_{\mathit{web}_1}&=\mathit{aim}_{\mathit{root}_{\mathit{web}_1}}=max(v_{1_{\mathit{web}}},v_{2_{\mathit{web}}},v_{3_{\mathit{web}}},v_{4_{\mathit{web}}}+v_{5_{\mathit{web}}}) \\
&=max(10.0,10.0,10.0,12.9)=12.9
\end{align*}

Let $\mathit{ap}$ denote the attack path captured in the HARM. We use $\mathit{ap}_1=\{\mathit{dns}_1,\mathit{web}_1,\mathit{app}_1,\mathit{db}_1\}$ and calculate $\mathit{aim}_{\mathit{ap}_1}$ in the following.
\begin{align*} 
\mathit{aim}_{\mathit{ap}_1}&={\mathit{aim}_{\mathit{dns}_1}}+{\mathit{aim}_{\mathit{web}_1}}+{\mathit{aim}_{\mathit{app}_1}}+{\mathit{aim}_{\mathit{db}_1}}\\
&=10.0+12.9+16.4+12.9=52.2
\end{align*}

Let $\mathit{AP}$ denote all the attack paths captured in the HARM, $\mathit{AIM}$ denote the attack impact in the network level. We calculate $\mathit{{AIM}_{befp}}$ for the network before patch in the following.
\begin{align*} 
\mathit{{AIM}_{befp}}&=\mathop{max}\limits_{\mathit{ap} \in \mathit{{AP}_{befp}}}{\mathit{aim}_\mathit{ap}}=52.2
\end{align*}

We show the security metrics for the HARMs before and after patch in Table~\ref{tb_results}. We denote the attack success probability as $\mathit{ASP}$, number of exploitable vulnerabilities as $\mathit{NoEV}$, number of attack paths as $\mathit{NoAP}$ and number of entry points as $\mathit{NoEP}$. From the results, we can conclude that patching the critical vulnerabilities increases the security of the example network.
\begin{table}[htb] \small
\caption{Security metrics for the example network.}
\label{tb_results}
\centering
\begin{tabular}{|c|c|c|c|c|c|}
\hline
\diagbox[width=0.825in]{HARM}{Metric} & $\mathit{AIM}$ & $\mathit{ASP}$ & $\mathit{NoEV}$ & $\mathit{NoAP}$ & $\mathit{NoEP}$\\
\hline
Before patch & 52.2 & 1.0 & 25 & 8 & 3\\
\hline
After patch & 42.2 & 0.265 & 11 & 4 & 2\\
\hline
\end{tabular}
\end{table}
 
\subsection{Construction of Availability Models}
\label{ava}

We assume a server consists of the hardware, OS and applications supporting the service. The server has both OS and application vulnerabilities to be patched. One or more of patches might require a reboot after installation to make the patches effective. There is no requirement of reboot between patches. At each patch period, application patches are performed at first and OS patches are performed immediately after application patches complete. Reboot occurs after both application and OS patches are finished in order to merge the reboot time. Other patch scenarios will be considered in our future work.

We make the following assumptions for the failure and recovery behaviors of the components in a server. Hardware may fail at anytime but will not fail during the patch period. Both OS and applications are subject to software failures. All patches are tested first in a pre-production environment before applying to production. So there are no software failures during the patch period. In addition, OS will not fail when it is ready to patch and applications will not fail when it is ready to patch. Other failure and recovery behaviors will be incorporated in our future work.

\begin{figure}[htb]
\centering
\includegraphics[width=3.5in]{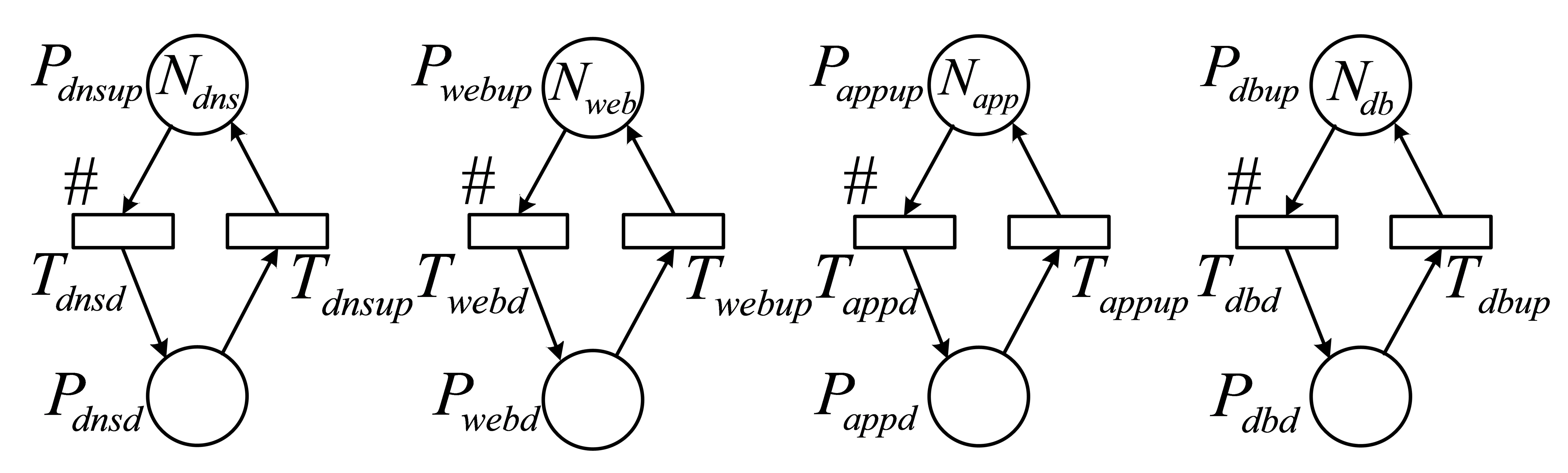}
\caption{SRN sub-models for the network.}
\label{fig_srn_net}
\end{figure}

\begin{figure*}[htb]
    \centering
    \begin{subfigure}{0.135\textwidth}
            \includegraphics[width=\textwidth]{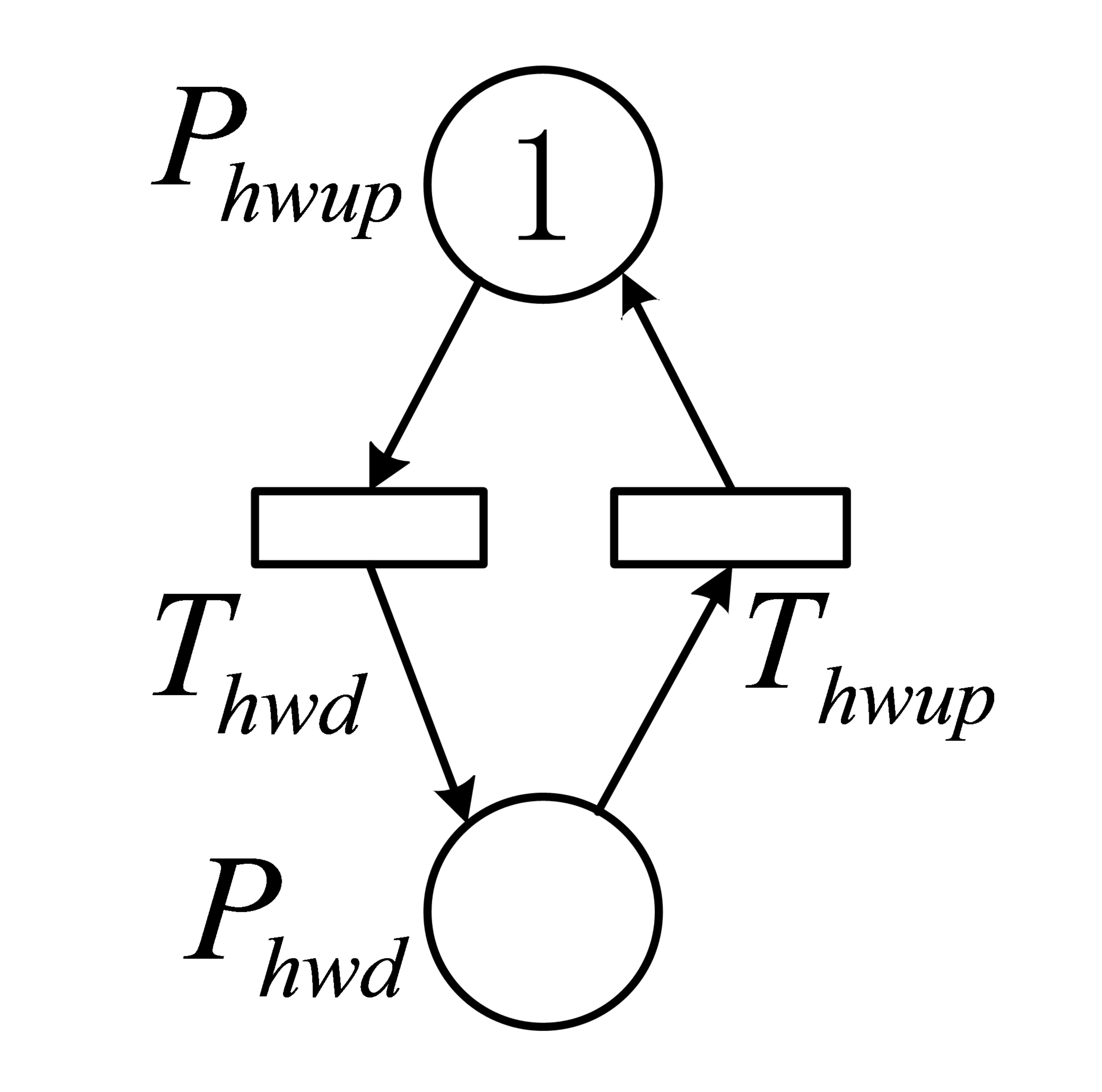}
            \caption{Hardware.}
    \end{subfigure}
    \begin{subfigure}{0.305\textwidth}
            \includegraphics[width=\textwidth]{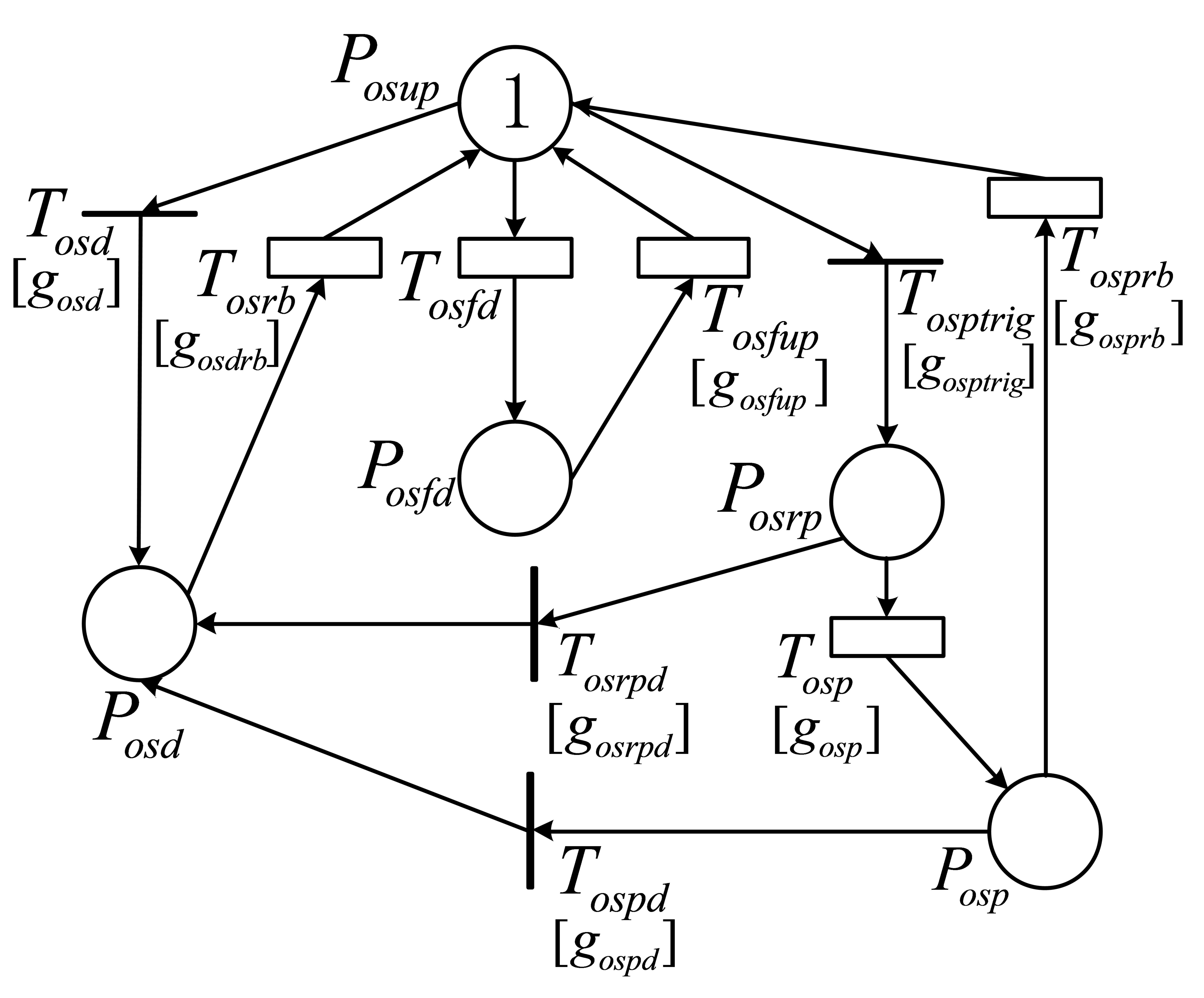}
            \caption{OS.}
    \end{subfigure}
    \begin{subfigure}{0.375\textwidth}
            \includegraphics[width=\textwidth]{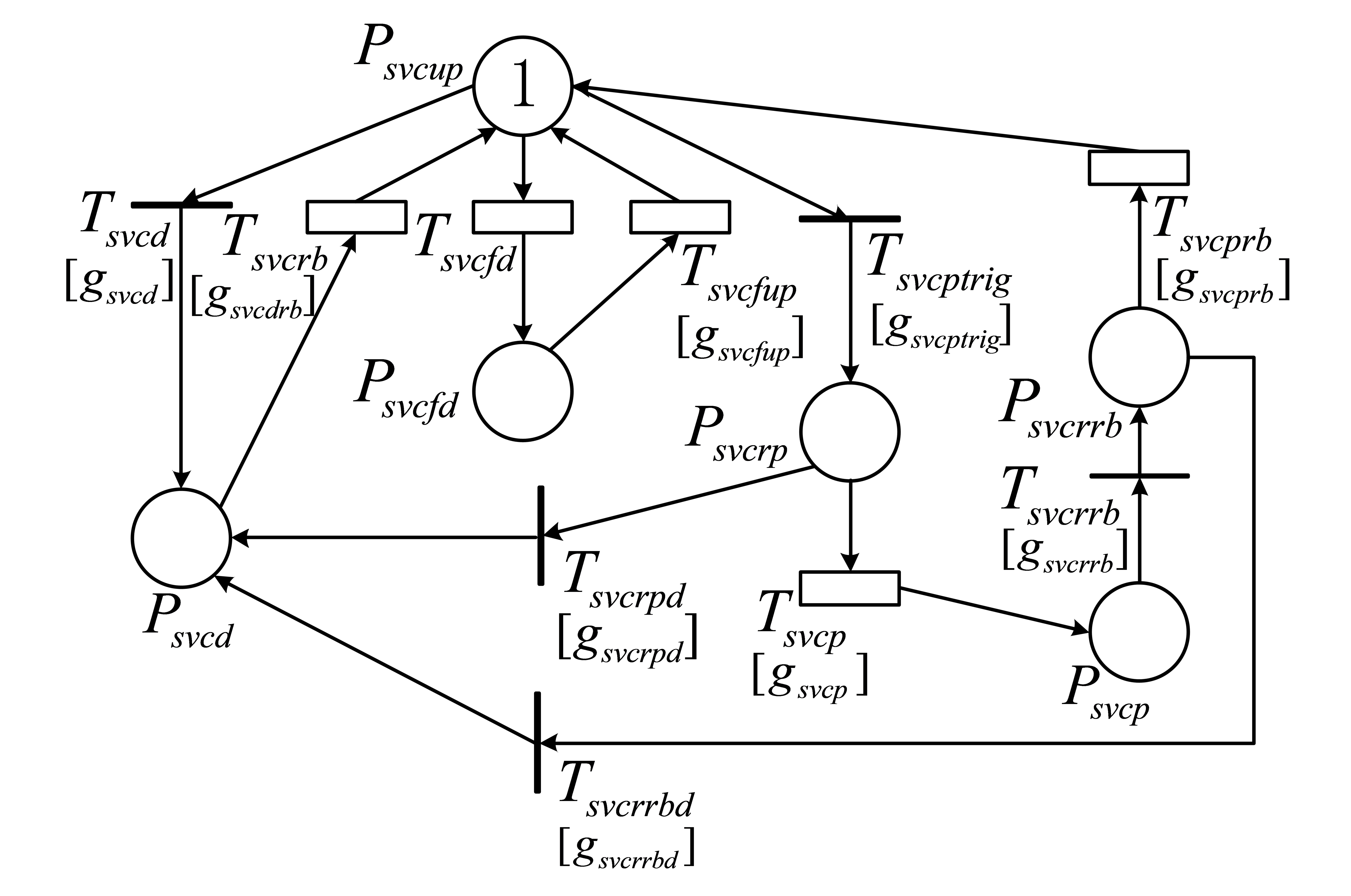}
            \caption{Service.}
    \end{subfigure}
    \begin{subfigure}{0.165\textwidth}
            \includegraphics[width=\textwidth]{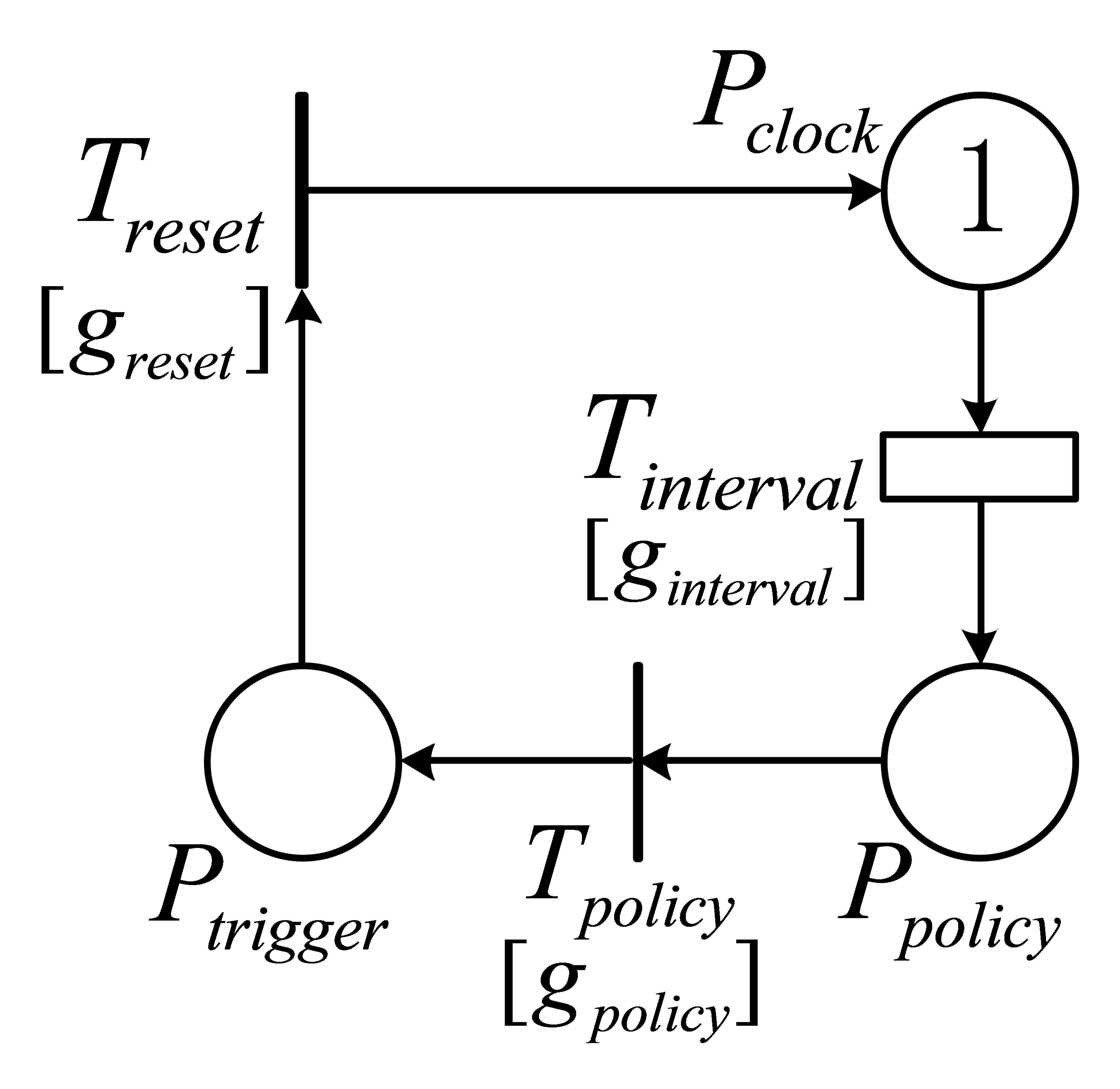}
            \caption{Patch clock.}
    \end{subfigure}
    \caption{SRN sub-models for a server.}
    \label{fig_srn_server}
\end{figure*}

We construct a hierarchical SRN model to evaluate the availability of the example network. The upper layer SRN sub-models shown in Figure~\ref{fig_srn_net} capture the dependencies of the servers in the network; the lower layer SRN sub-models shown in Figure~\ref{fig_srn_server} capture the dependencies of the components in a server. We describe the lower layer SRN sub-models in Section~\ref{sub_server} at first and the upper layer SRN sub-models in~\ref{sub_network} because the input parameters in the upper layer sub-models are calculated from the lower layer sub-models.

\subsubsection{SRN sub-models for a server}
\label{sub_server}

The SRN sub-models for a server consists of (a) hardware model, (b) OS model, (c) service model and (d) patch clock model. All the associated guard functions are defined in Table~\ref{tb_guard} where $\mathit{svc}$ denotes the service provided by a server.
\begin{table}[htb] \small
\caption{Guard functions in the SRN sub-models for a server.}
\label{tb_guard}
\centering
\begin{tabular}{|c|l|}
\hline
Guard & Definition\\
\hline
$g_{\mathit{osd}}$ & if (\#$P_{\mathit{hwd}}==1$) 1 else 0\\
\hline
$g_{\mathit{osdrb}}$ & if (\#$P_{\mathit{hwup}}==1$) 1 else 0\\
\hline
$g_{\mathit{osfup}}$ & if (\#$P_{\mathit{hwup}}==1$) 1 else 0\\
\hline
$g_{\mathit{osptrig}}$ & if (\#$P_{\mathit{svcp}}==1$) 1 else 0\\
\hline
$g_{\mathit{osp}}$ & if (\#$P_{\mathit{hwup}}==1$) 1 else 0\\
\hline
$g_{\mathit{osrpd}}$ & if (\#$P_{\mathit{hwd}}==1$) 1 else 0\\
\hline
$g_{\mathit{ospd}}$ & if (\#$P_{\mathit{hwd}}==1$) 1 else 0\\
\hline
$g_{\mathit{osprb}}$ & if (\#$P_{\mathit{hwup}}==1$) 1 else 0\\
\hline
$g_{\mathit{svcd}}$ & if (\#$P_{\mathit{hwd}}==1$ $||$ \#$P_{\mathit{osfd}}==1$) 1 else 0\\
\hline
$g_{\mathit{svcdrb}}$ & if (\#$P_{\mathit{hwup}}==1$ \&\& \#$P_{\mathit{osup}}==1$) 1 else 0\\
\hline
$g_{\mathit{svcfup}}$ & if (\#$P_{\mathit{hwup}}==1$ \&\& \#$P_{\mathit{osup}}==1$) 1 else 0\\
\hline
$g_{\mathit{svcptrig}}$ & if (\#$P_{\mathit{trigger}}==1$) 1 else 0\\
\hline
$g_{\mathit{svcp}}$ & if (\#$P_{\mathit{hwup}}==1$ \&\& \#$P_{\mathit{osup}}==1$) 1 else 0\\
\hline
$g_{\mathit{svcrpd}}$ & if (\#$P_{\mathit{hwd}}==1$ $||$ \#$P_{\mathit{osfd}}==1$) 1 else 0\\
\hline
$g_{\mathit{svcrrb}}$ & if (\#$P_{\mathit{osp}}==1$) 1 else 0\\
\hline
$g_{\mathit{svcrrbd}}$ & if (\#$P_{\mathit{hwd}}==1$ $||$ \#$P_{\mathit{osfd}}==1$) 1 else 0\\
\hline
$g_{\mathit{svcprb}}$ & if (\#$P_{\mathit{hwup}}==1$ \&\& \#$P_{\mathit{osup}}==1$) 1 else 0\\
\hline
\multirow{2}{*}{$g_{\mathit{interval}}$} & if (\#$P_{\mathit{svcup}}==1$ $||$ \#$P_{\mathit{svcd}}==1$ $||$ \#$P_{\mathit{svcfd}}==1$) 1\\
& else 0\\
\hline
$g_{\mathit{policy}}$ & if (\#$P_{\mathit{svcp}}==1$) 1 else 0\\
\hline
$g_{\mathit{reset}}$ & if (\#$P_{\mathit{osp}}==1$) 1 else 0\\
\hline
\end{tabular}
\end{table}

In Figure~\ref{fig_srn_server}(a), there is one token in place $P_{\mathit{hwup}}$ representing all typical hardware components of a server. Firing of $T_{\mathit{hwd}}$ represents the failure of any hardware component and firing of $T_{\mathit{hwup}}$ represents the recovery of the failed component. 

In Figure~\ref{fig_srn_server}(b), there is one token in place $P_{\mathit{osup}}$ representing the OS of a server. A token will be deposited in place $P_{\mathit{osd}}$ due to hardware failure by firing one of the three immediate transitions: $T_{\mathit{osd}}$, $T_{\mathit{osrpd}}$, $T_{\mathit{ospd}}$. The transition $T_{\mathit{osfd}}$ is fired when the OS fails. The immediate transition $T_{\mathit{osptrig}}$ is triggered when the application patch is finished. A token will be put in place $P_{\mathit{osp}}$ once OS patch is finished by firing $T_{\mathit{osp}}$.

In Figure~\ref{fig_srn_server}(c), there is one token in place $P_{\mathit{svcup}}$ representing the applications supporting the service of a server. A token will be deposited in place $P_{\mathit{svcd}}$ due to hardware or OS failure by firing one of the three immediate transitions: $T_{\mathit{svcd}}$, $T_{\mathit{svcrpd}}$, $T_{\mathit{svcrrbd}}$. The transition $T_{\mathit{svcfd}}$ is fired when any service application fails. The immediate transition $T_{\mathit{svcptrig}}$ is triggered when a token is in place $P_{\mathit{trigger}}$ of the patch clock model. A token will be put in place $P_{\mathit{svcp}}$ once application patch is finished by firing $T_{\mathit{svcp}}$. The immediate transition $T_{\mathit{svcrrb}}$ is triggered when the OS patch completes (i.e., a token in place $P_{\mathit{osp}}$).

In Figure~\ref{fig_srn_server}(d), there is one token in place $P_{\mathit{clock}}$ representing the patch clock of a server. The transition $T_{\mathit{interval}}$ is fired once per month. The immediate transition $T_{\mathit{policy}}$ is fired when the service is up. The immediate transition $T_{\mathit{reset}}$ is fired when the OS patch completes.

All transitions are assumed to have an exponentially distributed time duration. We estimate the failure and recovery rates of both hardware and software for each server based on~\cite{Kim2009PRDC}. We use the DNS server to demonstrate the input parameters of the SRN sub-models. From the NVD, we found one critical vulnerability in Microsoft DNS and two critical vulnerabilities in its Windows OS. We assume the vulnerability in the application needs 5 minutes to patch and the vulnerability in the OS needs 10 minutes to patch in average. The rate values of the DNS server are summarized in Table~\ref{tb_srn_input}.
\begin{table}[htb] \small
\caption{Input parameters of the SRN sub-models for the DNS server.}
\label{tb_srn_input}
\centering
\begin{tabular}{|c|c|c|c|}
\hline
Component & Transition & Parameter & Value\\
\hline
\multirow{2}{*}{Hardware} & Failure  & 1/$\lambda_{hw}$ & 87600 hours\\
\cline{2-4}
& Recovery & 1/$\mu_{hw}$ & 1 hour\\
\hline
\multirow{5}{*}{OS} & Failure  & 1/$\lambda_{os}$ & 1440 hours\\
\cline{2-4}
& Recovery & 1/$\mu_{os}$ & 1 hour\\
\cline{2-4}
& Patch & 1/$\alpha_{os}$ & 20 minutes\\
\cline{2-4}
& Reboot after patch & 1/$\beta_{os}$ & 10 minutes\\
\cline{2-4}
& Reboot after failure & 1/$\delta_{os}$ & 10 minutes\\
\hline
\multirow{5}{*}{DNS} & Failure & 1/$\lambda_{dns}$ & 336 hours\\
\cline{2-4}
& Recovery & 1/$\mu_{dns}$ & 30 minutes\\
\cline{2-4}
& Patch & 1/$\alpha_{dns}$ & 5 minutes\\
\cline{2-4}
& Reboot after patch & 1/$\beta_{dns}$ & 5 minutes\\
\cline{2-4}
& Reboot after failure & 1/$\delta_{dns}$ & 5 minutes\\
\hline
Patch clock & Time to patch & 1/$\tau_{p}$ & 720 hours\\
\hline
\end{tabular}
\end{table}

\subsubsection{SRN sub-models for the network}
\label{sub_network}

In order to analyze the availability affected by the patch schedule, we only consider the states and transitions caused by patch. We generate a two-state and two-transition CTMC for each server via the SRN sub-model, where the $up$ state represents the service is up and the $down$ state represents the service is down due to patch.

The firing rates of transitions $T_{\mathit{dnsd}}$, $T_{\mathit{webd}}$, $T_{\mathit{appd}}$ and $T_{\mathit{dbd}}$ are marking-dependent. Let $\lambda_{\mathit{eq}}^{\mathit{svc}}$ denote the patch rate, $\mu_{\mathit{eq}}^{\mathit{svc}}$ denote the recovery rate. The actual firing rate equals to $\lambda_{\mathit{eq}}^{\mathit{svc}}$\#${P_{\mathit{svcup}}}$. In the example network where $N_{\mathit{dns}}=1$, $N_{\mathit{web}}=2$,  $N_{\mathit{app}}=2$, $N_{\mathit{db}}=1$, the firing rates of transitions $T_{\mathit{dnsd}}$, $T_{\mathit{webd}}$, $T_{\mathit{appd}}$ and $T_{\mathit{dbd}}$ are $\lambda_{\mathit{eq}}^{\mathit{dns}}$, $2\lambda_{\mathit{eq}}^{\mathit{web}}$, $2\lambda_{\mathit{eq}}^{\mathit{app}}$, $\lambda_{\mathit{eq}}^{\mathit{db}}$, respectively.

We use the aggregation method in~\cite{Trivedi2013ASMBI} to compute the aggregated rates due to patch. Let $p_{\mathit{svcup}}$ denote the probability that the service is running, $p_{\mathit{svcpd}}$ denote the probability that the service is down due to patch, $p_{\mathit{svcprrb}}$ denote the probability that the service is ready to reboot after OS patch completes. We calculate $\lambda_{\mathit{eq}}^{\mathit{svc}}$ by Equation (\ref{eq_lambda}) and $\mu_{\mathit{eq}}^{\mathit{svc}}$ by Equation (\ref{eq_mu}). The probabilities in the equations can be calculated via the lower layer SRN sub-models for the server.
\begin{equation} \label{eq_lambda}
\lambda^{\mathit{svc}}_{\mathit{eq}}=\tau_{p}*p_{\mathit{svcup}}/p_{\mathit{svcup}}=\tau_{p}
\end{equation}
\begin{equation} \label{eq_mu}
\mu_{\mathit{eq}}^{\mathit{svc}}=\beta_{svc}*p_{\mathit{svcprrb}}/p_{\mathit{svcpd}}
\end{equation}

We show an example of calculating the patch rate and recovery rate for the DNS server by Equations (\ref{eq_lambda}) and (\ref{eq_mu}) in the following. $p_{\mathit{dnsprrb}}$ is the sum of the probabilities including all states that the DNS is down due to patch (i.e., OS and service are ready to patch and patched).
\begin{align*} 
\lambda^{\mathit{dns}}_{\mathit{eq}}&=\tau_{p}=1/720 \approx 0.001389\\
\mu_{\mathit{eq}}^{\mathit{dns}}&=\beta_{dns}*p_{\mathit{dnsprrb}}/p_{\mathit{dnspd}}\\
& \approx 12*0.00011563/0.00092506 \approx 1.49992
\end{align*}

We show the patch and recovery rates for all servers in Table~\ref{tb_aggregation_rates}. We also calculate the mean-time-to-patch (MTTP) using $1/\lambda_{\mathit{eq}}^{\mathit{svc}}$ and mean-time-to-recovery (MTTR) using $1/\mu_{\mathit{eq}}^{\mathit{svc}}$. All services have the same patch rate/MTTP as they are patched once per month. The application service has the lowest recovery rate (i.e., the longest MTTR) as the application server has more critical vulnerabilities to be patched.
\begin{table}[htb] \small
\caption{Aggregated values for the servers.}
\label{tb_aggregation_rates}
\centering
\begin{tabular}{|c|c|c|c|c|}
\hline
\multirow{2}{*}{\diagbox[width=0.85in]{Service}{Rate}} & MTTP & Patch & MTTR & Recovery\\
& (hour) & rate & (hour) & rate\\
\hline
DNS & 720 &0.00139 & 0.6667 & 1.49992\\
\hline
Web & 720 & 0.00139 & 0.5834 & 1.71420\\
\hline
Application & 720 & 0.00139 & 1.0001& 0.99995\\
\hline
Database & 720 & 0.00139 & 0.9167 & 1.09085\\
\hline
\end{tabular}
\end{table}

We choose the capacity oriented availability ($\mathit{COA}$) as the output measure. It can be obtained by computing the expected steady-state reward rate with the proper choices of reward rates for the model. We define the reward function of $\mathit{COA}$ in Table~\ref{tb_reward} where the reward rate is regarded as the number of running servers during patch divided by the total number of servers. We define four reward rates: 1, 0.83333 ($5/6$), 0.66667 ($4/6$) and 0. We use the upper layer SRN sub-models to calculate $\mathit{COA}$ which approximately equals to 0.99707. 
\begin{table*}[htb] \small
\caption{Reward function of $COA$ in the SRN sub-models for the network.}
\label{tb_reward}
\centering
\begin{tabular}{|c|l|}
\hline
Reward & Definition\\
\hline
\multirow{4}{*}{$\mathit{COA}$} & if (\#$P_{\mathit{dnsup}}==1$ \&\& \#$P_{\mathit{webup}}==2$ \&\& \#$P_{\mathit{appup}}==2$ \&\& \#$P_{\mathit{dbup}}==1$) 1\\
& else if (\#$P_{\mathit{dnsup}}==1$ \&\& \#$P_{\mathit{webup}}==1$ \&\& \#$P_{\mathit{appup}}==2$ \&\& \#$P_{\mathit{dbup}}==1$) 0.83333\\
& else if (\#$P_{\mathit{dnsup}}==1$ \&\& \#$P_{\mathit{webup}}==2$ \&\& \#$P_{\mathit{appup}}==1$ \&\& \#$P_{\mathit{dbup}}==1$) 0.83333\\
 & else if (\#$P_{\mathit{dnsup}}==1$ \&\& \#$P_{\mathit{webup}}==1$ \&\& \#$P_{\mathit{appup}}==1$ \&\& \#$P_{\mathit{dbup}}==1$) 0.66667 else 0\\
\hline
\end{tabular}
\end{table*}

\section{Numerical Analysis}
\label{analysis}

We can evaluate both the security and capacity oriented availability of a system under security patch. We assume there are several systems with different server redundancy. As an example, we deploy the active-active high availability cluster for each type of servers separately. We use the proposed approach to compare the security and capacity oriented availability for different designs and analyze the impact of security patch on these designs. We use identical servers for the same types of servers in different redundancy designs because this is commonly used in the enterprise networks. Heterogeneous redundant servers will be used in our future work.

\subsection{Comparison using two metrics}
\label{comparison}
\begin{figure*}[hbt]
    \centering
    \begin{subfigure}{0.31\textwidth}
            \includegraphics[width=\textwidth]{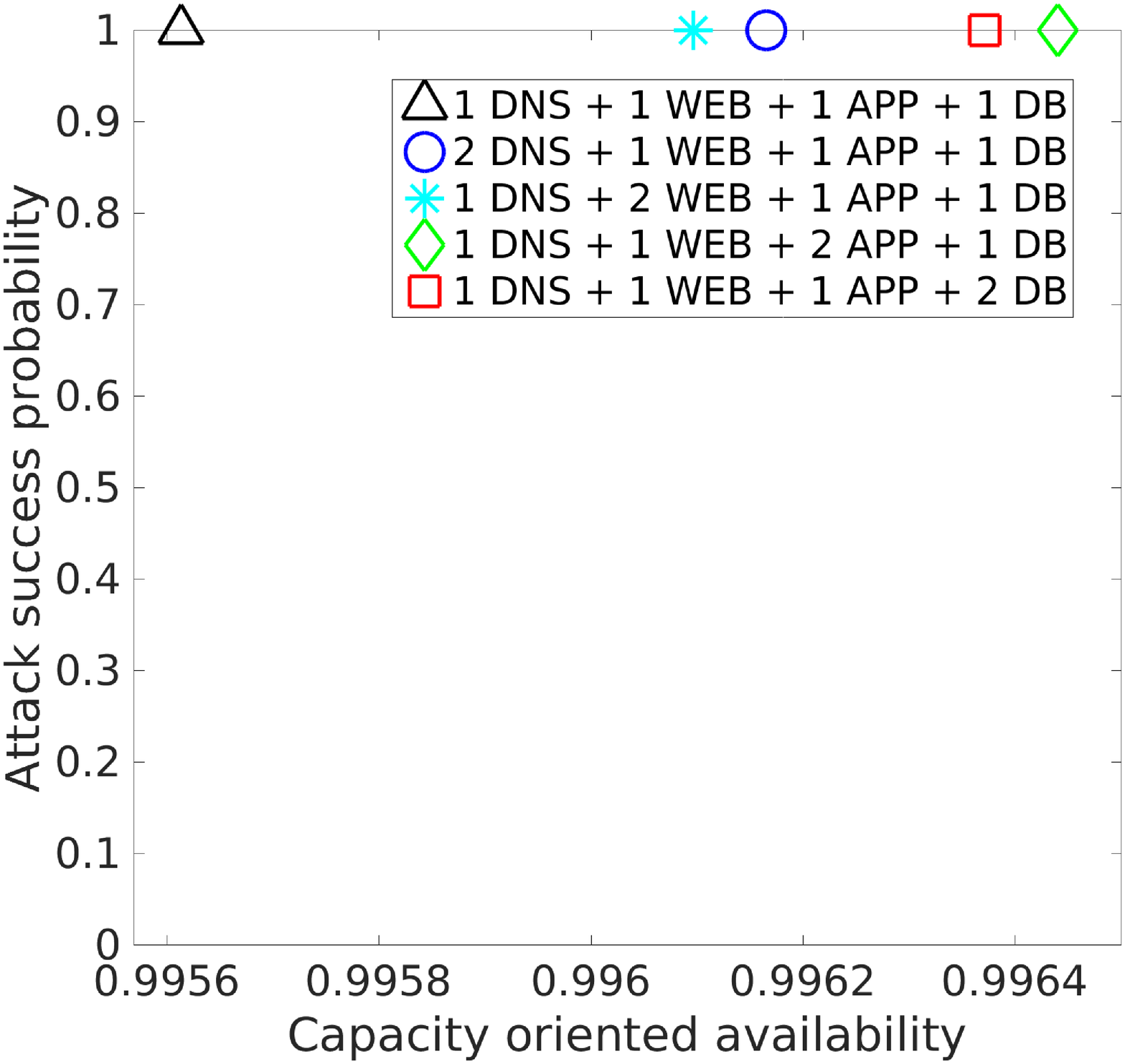}
            \caption{Before patch.}
    \end{subfigure}
    \begin{subfigure}{0.31\textwidth}
            \includegraphics[width=\textwidth]{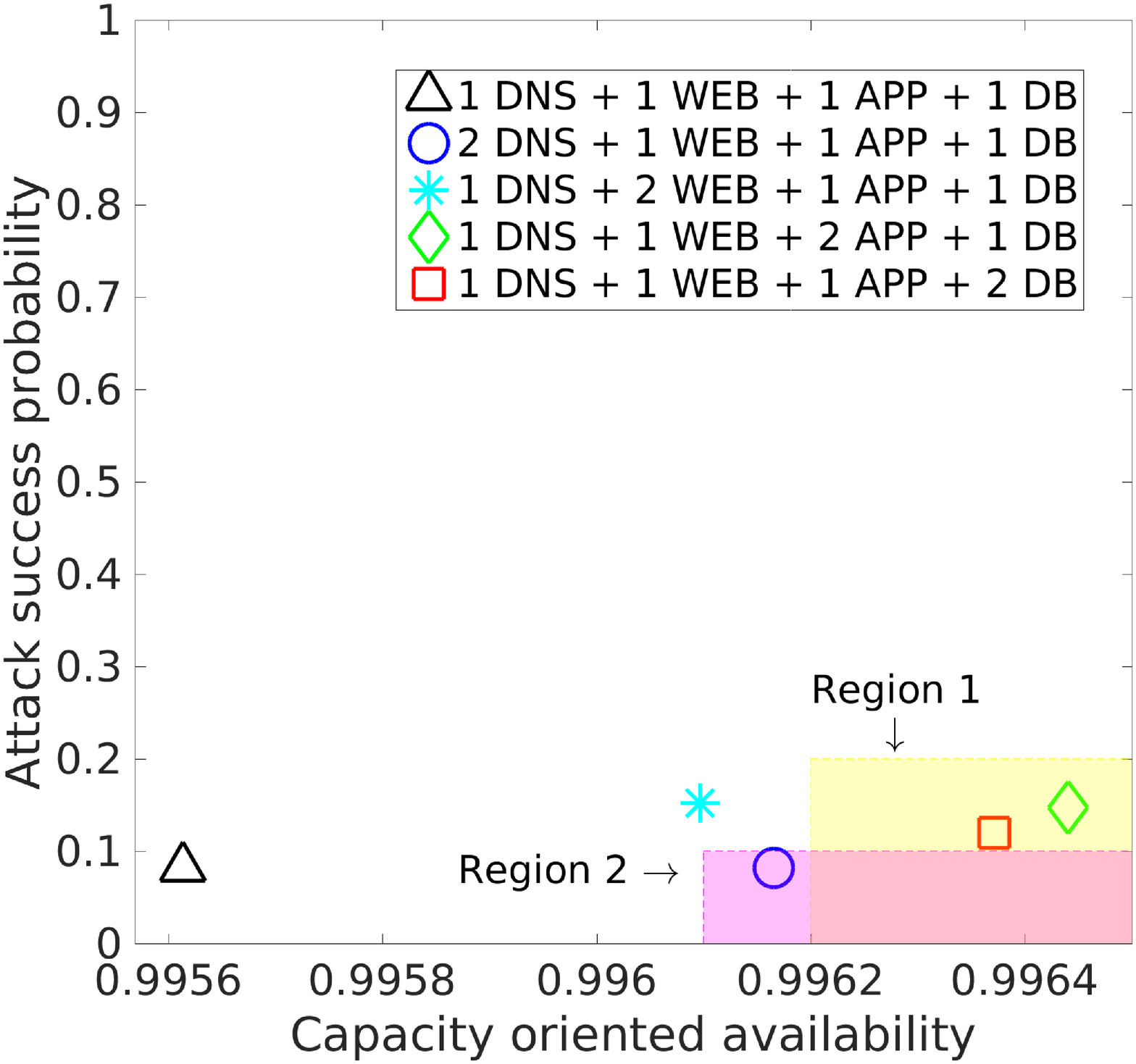}
            \caption{After patch.}
    \end{subfigure}
    \caption{Comparison of multiple redundancy designs using $ASP$ and $COA$.}
    \label{fig_scatter_asp}
\end{figure*}

We use the scatter plots to compare the results of one security metric and the availability metric. $\mathit{ASP}$ and $\mathit{COA}$ values before and after patch are shown in Figure~\ref{fig_scatter_asp}(a) and~\ref{fig_scatter_asp}(b). In Figure~\ref{fig_scatter_asp}(a), before patch, all redundancy designs have the maximum value of attack success probability. This is an extreme situation as each server initially has at least one vulnerability with the maximum value of attack success probability. When calculating the value for the server in an AT, each server also has the maximum probability value thus causing the maximum probability value in the system level. In Figure~\ref{fig_scatter_asp}(b), security patch decreases $\mathit{ASP}$ for all redundancy designs as all critical vulnerabilities are fixed. The increasing level of redundancy increases $\mathit{COA}$ but reduces $\mathit{ASP}$. In the fourth design, $\mathit{COA}$ is higher than the values in other designs as the application service has the lowest recovery rate shown in Table~\ref{tb_aggregation_rates}. In addition, the first and second designs have the same $\mathit{ASP}$ because the DNS server has no exploitable vulnerabilities after patch. All other designs have higher $\mathit{ASP}$ than the first non-redundancy design.

We define a function shown in Equation (\ref{eq_func1}) which compares values of metrics with the upper/lower bounds of metrics defined by the administrator. Let $\phi$ denote the upper bound of $\mathit{ASP}$, $\psi$ denote the lower bound of $\mathit{COA}$. The output of the function is either 1, indicating the design satisfies the requirements, or 0, indicating the design does not satisfy the requirements.
\begin{equation} \label{eq_func1}
f(\mathit{ASP}, \mathit{COA})=
\left\{
\begin{array}{ll}
1,& \text{if } \mathit{ASP} <= \phi \text{ and } \mathit{COA} >= \psi\\
0,& \text{if } \mathit{ASP} > \phi \text{ or } \mathit{COA} < \psi
\end{array}
\right.
\end{equation}

We assume two upper bounds of $\mathit{ASP}$ and two lower bounds of $\mathit{COA}$ for the designs after security patch are defined by an administrator. Design choices satisfying both security and availability are shown as follows.
\begin{enumerate}
\item $\phi = 0.2$ and $\psi = 0.9962$ (region 1): 1 DNS + 1 WEB + 2 APP + 1 DB; 1 DNS + 1 WEB + 1 APP + 2 DB.
\item $\phi = 0.1$ and $\psi = 0.9961$ (region 2): 2 DNS + 1 WEB + 1 APP + 1 DB.
\end{enumerate}

\subsection{Comparison using multiple security and availability metrics}

\begin{figure*}[hbt]
    \centering
    \begin{subfigure}{0.5\textwidth}
            \includegraphics[width=\textwidth]{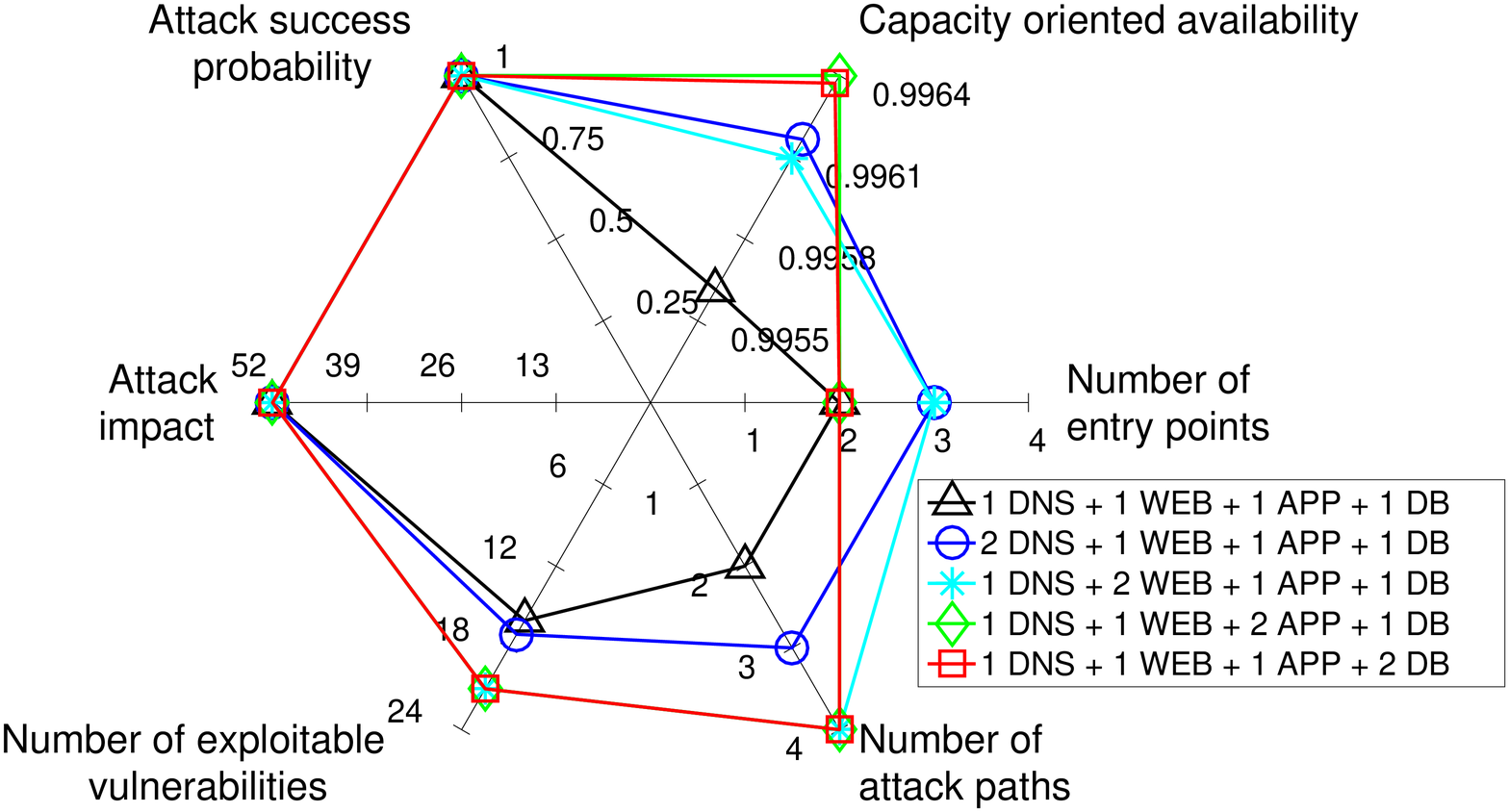}
            \caption{Before patch.}
    \end{subfigure}
    \begin{subfigure}{0.47\textwidth}
            \includegraphics[width=\textwidth]{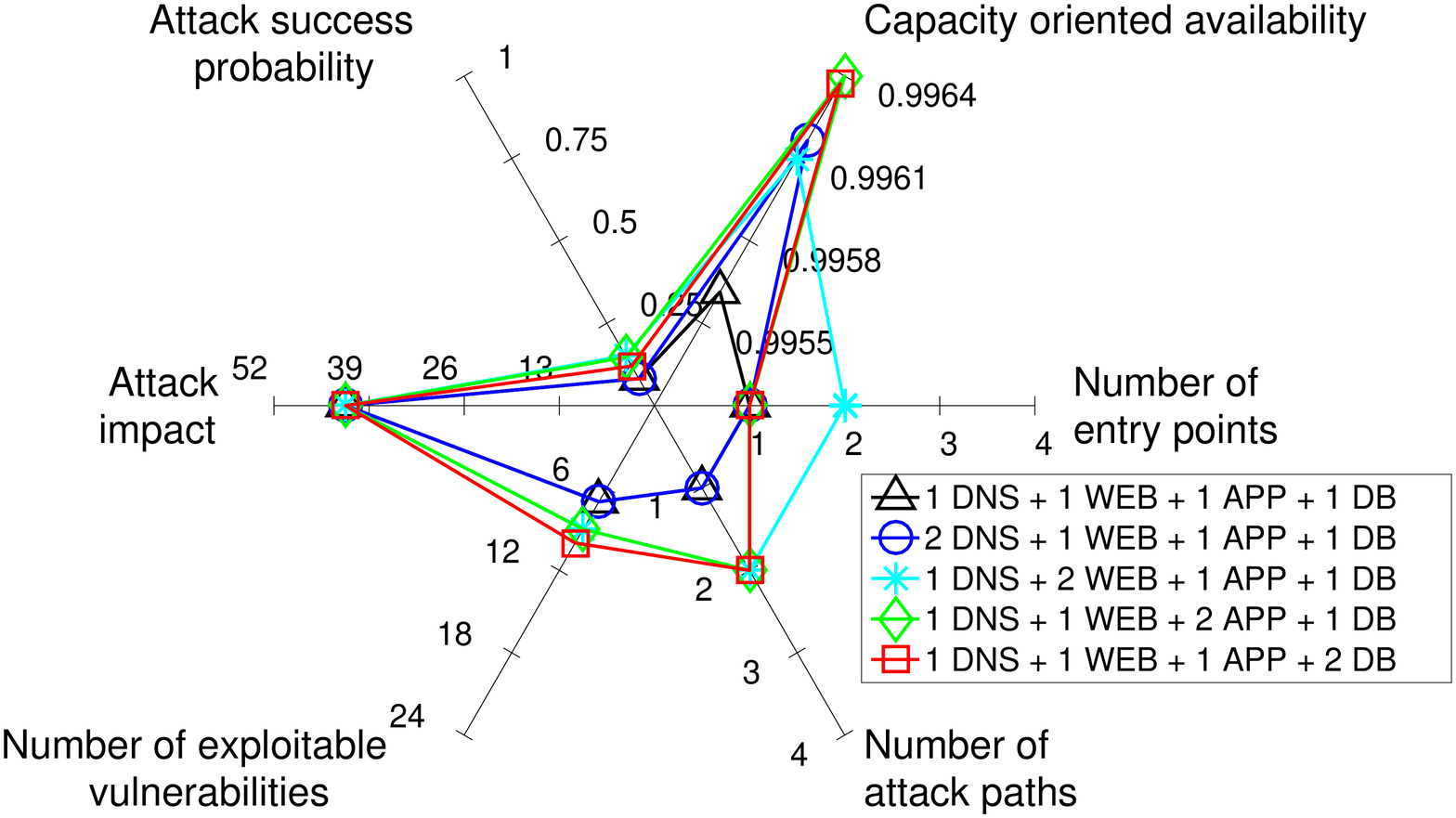}
            \caption{After patch.}
    \end{subfigure}
    \caption{Comparison of multiple redundancy designs using various metrics.}
    \label{fig_radar}
\end{figure*}

We can also compare the results of multiple metrics via the radar charts. The output values of six metrics before and after patch are shown in Figure~\ref{fig_radar}(a) and~\ref{fig_radar}(b). 

In Figure~\ref{fig_radar}(a), $\mathit{AIM}$ does not change in all the design choices. As shown in Section~\ref{sec}, $\mathit{AIM}$ is the maximum impact value among all the values in the attack path level. Before patch, the maximum impact value in the attack path level is calculated from the longest path including DNS, web, application and database servers. Each redundancy design has the same longest attack path which causes the same $\mathit{AIM}$. The increasing level of redundancy increases $\mathit{COA}$ but reduces $\mathit{NoEV}$ and $\mathit{NoAP}$. In addition, the fourth and fifth designs have the same $\mathit{NoEP}$ as the first non-redundancy design because the redundant server is not the entry point.

In Figure~\ref{fig_radar}(b), the values of all security metrics drop after the patch. As the DNS server has no exploitable vulnerability after the critical vulnerability is patched, each attack path only includes web, application and database servers. Therefore $\mathit{AIM}$ does not change for all designs due to the same attack path. In addition, the first and second designs have the same $\mathit{NoAP}$ and $\mathit{NoEV}$ because the DNS server is excluded from the attack path. All other designs have higher $\mathit{NoAP}$ and $\mathit{NoEV}$ than the first non-redundancy design. Only the third design has higher $\mathit{NoEP}$ as the redundant web servers still have exploitable vulnerabilities.

We define a function shown in Equation (\ref{eq_func2}) which compares values of metrics with the upper/lower bounds of metrics defined by an administrator. As $\mathit{AIM}$ is the same in all design choices, we assume the administrator defines the upper bounds for $\mathit{ASP}$, $\mathit{NoEV}$, $\mathit{NoAP}$ and $\mathit{NoEP}$ and the lower bound for $\mathit{COA}$. Let $\xi$ denote the upper bound of $\mathit{NoEV}$, $\omega$ denote the upper bound of $\mathit{NoAP}$, $\kappa$ denote the upper bound of $\mathit{NoEP}$.
\begin{multline} \label{eq_func2}
f(\mathit{ASP}, \mathit{NoEV}, \mathit{NoAP}, \mathit{NoEP}, \mathit{COA})=\\
\left\{
\begin{array}{ll}
1,&\text{if } \mathit{ASP} <= \phi \text{ and }\\
&\mathit{NoEV} <= \xi \text{ and }\\
&\mathit{NoAP} <= \omega \text{ and }\\
&\mathit{NoEP} <= \kappa \text{ and }\\
&\mathit{COA} >= \psi\\
0,& \text{if } \mathit{ASP} > \phi \text{ or }\\
&\mathit{NoEV} > \xi \text{ or }\\
&\mathit{NoAP} > \omega \text{ or }\\
&\mathit{NoEP} > \kappa \text{ or }\\
&\mathit{COA} < \psi
\end{array}
\right.
\end{multline}

We assume the administrator defines two upper bounds of $\mathit{ASP}$, $\mathit{NoEV}$, $\mathit{NoAP}$ and $\mathit{NoEP}$ and two lower bounds of $\mathit{COA}$ for the designs after security patch. The design choices which satisfy both security and availability are shown as follows.
\begin{enumerate}
\item $\phi = 0.2$ and $\xi = 9$ and $\omega = 2$ and $\kappa = 1$ and $\psi = 0.9962$: 1 DNS + 1 WEB + 2 APP + 1 DB.
\item $\phi = 0.1$ and $\xi = 7$ and $\omega = 1$ and $\kappa = 1$ and $\psi = 0.9961$: 2 DNS + 1 WEB + 1 APP + 1 DB.
\end{enumerate}

\subsection{Summary}

From the above numerical analysis, there is a balance between the security and capacity oriented availability of redundancy designs under security patch. As the redundancy designs may bring negative effect on security in terms of attack success probability, number of exploitable vulnerabilities, number of attack paths and number of entry points, high security and availability cannot be achieved at the same time. We have several observations that will facilitate an administrator's decision making on redundancy designs under security patch.
\begin{itemize}
\item Increasing the redundancy of the server with lowest recovery rate (i.e., longest mean time to patch critical vulnerabilities and reboot) has better improvement on the capacity oriented availability of the network under security patch;
\item The redundant servers with no exploitable vulnerabilities after patch do not decrease security while improve the system availability.
\end{itemize}

\section{Limitations and Extensions}
\label{limit}

We plan to complete the following extensions in our future work.

\textbf{Systems:} in the current analysis, we use small-scale enterprise networks with identical redundant servers. Larger number of servers and heterogeneous redundancy will be introduced and evaluated. Besides, we will apply the proposed approach to real networks to analyze the impact of security patch on security and availability when introducing redundant servers and compare different design choices.

\textbf{SRN models:} there are several limitations about the current SRN model. The SRN model only works for specific patch scenarios as mentioned in the assumptions. Sometimes a server might only have OS vulnerabilities or application vulnerabilities to be patched. Some patches might not need to reboot the application or the OS. We will re-design the SRN model to accommodate the complex scenarios in the real world. 

\textbf{User oriented performance:} the current work does not consider the performance of the redundancy design under client requests. We can use queuing network to model the arrival and processing of client requests and compute performance measures (e.g., mean response time, mean waiting time).

\textbf{Other metrics:} more metrics can be used for the model-based evaluation. For example, the operational cost can be added as the output measure of the SRN model to compare the costs of different redundancy designs. Besides, we will use economic metrics to help the administrators make decisions on redundancy designs (e.g., gain of high availability versus cost of redundancy; loss of successful attacks versus cost of security patch).

\textbf{Patch schedule}: we use monthly patch schedule in the analysis. Other patch schedules can also be used in the evaluation. Impact of different patch schedules on security and availability can be compared.

\section{Conclusions}
\label{conclusion}

Security patch can improve the security of enterprise networks but introduce downtime. Redundancy designs can be used to improve the availability under security patch. However, introducing redundant servers also increases the attack surface. It is important to find the balance between the security and availability affected by the security patch. 

In this paper, we have shown both security and availability models to assess the security and capacity oriented availability of multiple design choices for the server redundancy under the security patch. We have carried out model-based evaluation using five design choices with different number of redundant servers. We have used scatter plots to compare two metrics and radar charts to compare multiple metrics. The results have shown most redundancy designs increase capacity oriented availability but decrease security under security patch. The exception is that if the redundant server has no exploitable vulnerability after patch, this redundancy design will not decrease security while improving the system availability. Besides, we have defined functions which compare values of metrics with the upper/lower bounds to find design choices that satisfy both security and availability requirements.

\bibliographystyle{IEEEtran}
\bibliography{Availability,IoTsecurity}

\end{document}